\PassOptionsToPackage{table}{xcolor} 
\documentclass[sigconf]{acmart}

\definecolor{graylight}{rgb}{0.85,0.85,0.85}
\definecolor{grayextralight}{rgb}{0.95,0.95,0.95}

\usepackage{colortbl} 

\usepackage{graphicx}
\usepackage{hyperref} 
\usepackage{hyperxmp} 
\usepackage{float}
\usepackage{tabularx}
\usepackage{multirow} 
\usepackage{array} 
\usepackage{geometry} 
\usepackage{multicol} 
\usepackage{geometry}
\usepackage{tcolorbox} 

\newcommand{\interviewquote}[2]{{\small\faComments\,#1:\textit{``#2"}}}

\def\BibTeX{{\rm B\kern-.05em{\sc i\kern-.025em b}\kern-.08emT\kern-.1667em\lower.7ex\hbox{E}\kern-.125emX}}
    
\usepackage{pgfplots}
\pgfplotsset{width=10cm,compat=1.9}

\usepackage{stackengine}
\usepackage{verbatim}
\usepackage{xspace}

\usepackage{xcolor}
\definecolor{darkgreen}{rgb}{0.05,0.5,0.05}

\newcommand{\added}[1]{{#1}\xspace}

\renewenvironment{quote}{%
   \list{}{%
     \leftmargin\parindent
     \rightmargin0cm
   }
   \item\relax
}
{\endlist}

\usepackage{fontawesome5}

\usepackage{float}
\newcommand\revision[1]{}

\begin{document}

%
\title{Dear Diary: A randomized controlled trial of Generative AI coding tools in the workplace}

\author{Jenna Butler}
\affiliation{%
  \institution{Microsoft}
  \city{Redmond}
  \state{Washington}
  \country{USA}
}
\email{jennbu@microsoft.com}

\author{Jina Suh}
\affiliation{%
  \institution{Microsoft}
  \city{Redmond}
  \state{Washington}
  \country{USA}
}
\email{jinsuh@microsoft.com}

\author{Sankeerti Haniyur}
\affiliation{%
  \institution{Microsoft}
  \city{Redmond}
  \state{Washington}
  \country{USA}
}
\email{sahaniyur@microsoft.com}

\author{Constance Hadley}
\affiliation{%
  \institution{Institute for Work Life} 
  \city{Boston}
  \state{Massachusetts}
  \country{USA}
}
\email{cnhadley@institutelifework.org}

%
\renewcommand{\shortauthors}{Butler et al.}

%
\begin{abstract}
Generative AI coding tools are relatively new, and their impact on developers extends beyond traditional coding metrics, influencing beliefs about work and developers' roles in the workplace. This study aims to illuminate developers' pre-existing beliefs about generative AI tools, their self-perceptions, and how regular use of these tools may alter these beliefs. Using a mixed-methods approach, including surveys, a randomized controlled trial, and a three-week diary study, we explored the real-world application of generative AI tools within a large multinational software company. Our findings reveal that the introduction and sustained use of generative AI coding tools significantly increases developers' perceptions of these tools as both useful and enjoyable. However, developers' views on the trustworthiness of AI-generated code remained unchanged. We also discovered unexpected uses of these tools, such as replacing web searches and fostering creative ideation. Additionally, 84\% of participants reported positive changes in their daily work practices, and 66\% noted shifts in their feelings about their work, ranging from increased enthusiasm to heightened awareness of the need to stay current with technological advances. This research provides both qualitative and quantitative insights into the evolving role of generative AI in software development and offers practical recommendations for maximizing the benefits of this emerging technology—particularly in balancing the productivity gains from AI-generated code with the need for increased scrutiny and critical evaluation of its outputs.
\end{abstract}

%

\maketitle


\section{Introduction}
Generative AI coding tools, such as Github Copilot, have been touted as being able to cut programming tasks by 55\% (\cite{copilot_eirini}), an incredible finding obtained during a lab experiment looking at writing an HTTPS server. Shortly after release of Github Copilot, studies were conducted attempting to answer the many questions popping up with the release of a totally new paradigm of tools - did they reduce overall time? Did they help junior people up-level? Do people write more code with these tools? However, most of these studies were lab studies and looking at usability and quantitative observations (lines of code, time taken, etc). 

Developer tools have the potential to impact engineers in their real world jobs in existing codebases and, perhaps even more importantly, in how engineers actually feel about these tools and their work. 
For example, prior research has found that beliefs about a tool significantly influence its adoption (\cite{beliefs_moeinedin}, \cite{beliefs_xiao}, \cite{beliefs_agarwal} and \cite{beliefs_vishwanath}). As such, we set out to understand developers beliefs about Generative AI coding tools, how they used them in a real world setting, and how the use of such tools changed their beliefs about the tools or themselves as engineers.

To understand the changes in developers' early beliefs that can be isolated to the use of Generative AI coding tools, we conducted a randomized controlled trial of Github Copilot in a real world workplace. 
We enrolled over 200 engineers at a major software organization and surveyed them to learn their pre-existing beliefs about Generative AI tools, as well as their beliefs about themselves as an engineer in their workplace. We then used blocked randomization to group them into a control group and a treatment group and gave the treatment group access to Github Copilot for 3 weeks. Both prior to usage and during, we measured telemetry such as lines of code written, hours spent coding, and total PRs created and submitted. After the three weeks, we re-surveyed the total population to look at how the use they used Github Copilot and how this use changed their beliefs and values about tools and their own place in the engineering workforce. With this data, we answered the following research questions:

\begin{itemize}
    \item \textbf{RQ1}: What pre-existing \textbf{values and beliefs} do developers have about GenerativeAI coding tools?
    \item \textbf{RQ2}: How do pre-existing \textbf{values and beliefs} change after regular use of GenerativeAI code tools in their daily work? 
    \item \textbf{RQ4}: What are the actual \textbf{use cases} and \textbf{challenges} of using GenerativeAI in real work?
    \item \textbf{RQ3}: What is the \textbf{impact} of using GenerativeAI in developer's daily work, and how does it \textbf{change their work and their place in it}?
\end{itemize}

Our findings indicate that the use of generative AI tools in the workplace leads to a statistically significant increase in the belief that these tools are useful. Additionally, developers who utilize generative AI tools report positive changes in both how they work and how they perceive their work, as these tools reduce mundane tasks and position developers at the cutting edge of technology.

This paper contributes one of the first randomized controlled trials of GitHub Copilot in a real-world work environment, examining not only the impacts on quantitative coding data but also the effects of these tools on developers' beliefs and values. We provide insights into the preconceived notions developers hold about these tools — often before ever using them — and explore if continued use of these tools changes or reinforces preconceived beliefs. Finally, we highlight the real-world use cases developers find for these tools, emphasizing the importance of leveraging their benefits while maintaining a critical perspective on the reliability of large language model outputs.


\section{Background and Related Work}
Much research has existed how beliefs about a tool influence adoption, and while generative AI programming assistants are faily new, the body of research on them is growing quickly. Here we will explore prior work on the impact of beliefs on tool adoption in general as well as specifically in the software world and early studies of Github Copilot and other GenAI programming tools.

\subsection{Impact of Beliefs on Tool Adoption}
Prior studies have found that beliefs across many areas impact tool adoption, such as perceived usefulness \cite{beliefs_moeinedin}, \cite{beliefs_mingolla}, \cite{beliefs_riemenschneider} and social dynamics, such as what others think about the tool \cite{beliefs_riemenschneider}, \cite{beliefs_ritter}, \cite{beliefs_xiao}, \cite{beliefs_swan}. Moeindedin and Mingolla \textit{et al.} both found that perceived usefulness is a predictor of adoption, with Moeindedin finding that perceived usefulness was a better predictor than perceived ease of use for an electronic tool for cancer diagnosis \cite{beliefs_moeinedin}. Mingolla \textit{et al.} looked at cattle farmer's adoption of diagnostic tools and found that perceived usefuless as well as economical value were predictors of adoption. Not just tools, but also new methodologies are impacted by what this perception of usefulness. Riemenschneider \textit{et al.} found that developer acceptance of new methodologies was influenced by usefulness perception \cite{beliefs_riemenschneider}.

Another common finding is that the thoughts and perceptions of others influences adoption. In addition to finding usefulness perception impacts, Riemenschneider \textit{et al.} found that what others (such as coworkers and supervisors) thought of a tool influenced adoption of a new methodology \cite{beliefs_riemenschneider}. Again, Ritter also found this, finding that the opinion of vetrinarians and other farmers was a key influence in the adoption of new practices in farming \cite{beliefs_ritter}. In the software field, Xiao \textit{et al.} found that when it comes to adoption of new security tools, what other colleagues think as well as the opinions of trusted online personalities has an effect \cite{beliefs_xiao}. What an individual thinks about a tool impacts adoption for themselves as well, as Swan and Newell found that a managers belief system plays a role in how they decide what tools to adopt \cite{beliefs_swan}. In the software industry, more than just perceived usefulness, personal beliefs and the beliefs of those around you has been found to impact adoption. Agarwal and Prasad found that prior knowledge, org tenure, training experience and perceived job insecurity can also impact beliefs \cite{beliefs_agarwal}. Vishwanath also found that how a tool is framed impacts adoption \cite{beliefs_vishwanath}. Adoption of ChatGPT as a developer tool has been impacted by expectations around performance and security \cite{beliefs_ge}. When people do decide to adopt a new tool, early adoptors are viewed with a higher social standing, compared to those who adopt the tool later \cite{beliefs_lamba}.

\subsection{Generative AI Coding Tools}
Github published a blog post early on, looking at the impact of AI on developer productivity through their tool, Github Copilot \cite{copilot_eirini}. The authors conducted a lab experiment in Spring 2022 where 95 professional programmers were recruited to participate and attempt to write an HTTP server in JavaScript either with Github Copilot or without.  While there was not a statistically significant difference in the completion rate between the two groups, they did find a 55.8\% reduction in programming time for the treatment group with Github Copilot \cite{copilot_eirini}. Other studies have continued to try and measure the impact of GenAI coding tools. Chatterjee \textit{et al.} also looked at the productivity impacts of Github Copilot - this time with programming tasks completed by programmers at ANZ bank. They found developers with Copilot spent statistically signficantly less time working on these programming tasks, with an average productivity boost of 42.36\% \cite{genai_chatterjee}. Another study found that PRs reviewed by Github Copilot Pull Rest tool had an average review time 19.3 hours less than those PRs not reviewed by Copilot \cite{genai_xiao}.

The productivity gains of using Github Copilot are not always as straightforward. Pandey \textit{et al.} looking at 15 development tasks conducted in a real world codebase using Copilot found a 50\% time savings in code documentation and autocompletion \cite{genai_pandey}. However, they also found Copilot struggled with large functions and complex tasks, which caused them to lower their projected time savings to 33-36\%. Imagi also found this nuance when they compared using Github Copilot as a pair programming partner to using a human \cite{genai_imai}. They found that while Copilot increases how many lines of code are added, the quality isn't as high as when two humans code together. Another study generated 1,689 programs and found that approximately 40\% of them contained security vulnerabilities \cite{genai_pearce}. Additionally, Arghavan \text{et al.} compared GenAI solutions to fundamental coding problems (such as sorting and data structure implementations) and found that humans still solved the problems more often than GenAI, although the bugs GenAI introduced were easier to solve than bugs created by humans \cite{genai_arghavan}. 

Since these tools are still quite new, researchers continue to try to understand how developers use them in their daily lives. Zhang \textit{et al.} analyzed 169 Stack Overflow posts and 655 Github Discussions to learn what programming languages, IDEs and technologies were being used with Copilot and what benefits, limitations and challenges were being reported \cite{genai_beiqi}. They found it was mostly being used with JavaScript and Python in VSCode and that people were using it for data processing and found the code generation useful, but found lack of integration difficult. Suad looked at patterns of conversations developers tend to have with LLMs and found developers use these tools for both traditional software tasks (such as learning about frameworks and programming langauges) and more general tasks, like writing assistance and general queries \cite{genai_suad}. Indeed, developers report liking using GenAI assistance for writing tests and natural-language artifacts, since these are often tasks they don't typically enjoy in their job \cite{genai_sergeyuk}.

Looking ahead, some have predicted that with GenAI coding tools now available, the developer role with shift more to reviewing code than writing it \cite{genai_bird}. That said, France believes the shift will be somewhat slow - not a rapid change where developers are replaced by AI, but a shift where developers must adapt and learn this new way of working \cite{genai_france}.

\section{Methodology}
\label{sec:methodology}

Our study goals are to understand developers' existing beliefs about Generative AI coding tools and to observe how the use of such tools change their beliefs about the tools or themselves as engineers over time.
To do this, we conducted a randomized controlled trial with surveys, diary entries, and telemetry measurement between those that used Github Copilot for the first time and those that did not use the tool at all. Our study was primarily inductive, as very little was known at the time of the study about how GenAI coding tools may actually transform developer beliefs. 
In this section, we describe the study, starting with the organizational setting in which the study was conducted, followed by a description of the design.

\subsection{Research Setting}
This study took place in a large multi-country software company with engineers from around the world. These engineers work in one of the largest code-bases in the world, comprised of over 1B lines of code, that ships to multiple endpoints including mobile, web and desktop, and has files ranging from 35 years old up to brand new. These factors make this a difficult and complex system and software engineers working in it need to be especially proficient.

Github Copilot was available in this organization starting in summer 2022, but did not yet have wide spread adoption and was an optional addition to a developer's programming environment. This meant we had a small window of opportunity to do a randomized study of engineers before they had ever interacted with Github Copilot. With Github Copilot being said to make developers 55\% more productive in a lab setting, we were eager to see the impact of Github Copilot in a real world workplace. Beyond productivity impacts, we were especially interested in how it could potentially impact the developers themselves. The addition of any tool into a workplace is inherently a sociotechnical one, and we wanted to see not just how it impacted standard productivity measures (such as lines of code, number of pull requests (PRs), etc) but how it influenced developers beliefs about their workplaces and themselves as engineers. We wondered if developers would feel empowered and excited about a new tool, or if they would feel dejected and fear for their jobs after seeing the abilities of Github Copilot. We designed this study to investigate these conflicting hypotheses.

\subsection{Overall Design}
The study was designed as a randomized controlled trial that examined the impact of Github Copilot. 
The population under study involved software engineers from a large software organization, on different teams, in different countries, and with different languages and tools. The overall procedure of the study is presented briefly here, and we go into further details on each element of the study below. First, we recruited engineers to participate in a randomized controlled diary study of the new Github Copilot and consented participants filled out an intake survey about their current knowledge of and beliefs around Generative AI coding tools. They were then separated into 3 groups: those who had already been using Github Copilot and would remain using it; those who had not and now would be given access (treatment); and those who had not and would continue to not use it (control). We provided instructions to each group to either continue working as normal, on how to install and use Github Copilot, or to not use any generative AI coding tools for the next 3 weeks, and sent out a daily diary for participants to fill out about their day, their perceived productivity, and their use of the tool (if using it). At the end of three weeks, sent a closing survey which asked the same beliefs questions as before, and asked for any closing thoughts, as well as self-reported compliance. We also collected telemetry (such as lines of code written, PRs sent, hours spent coding, etc.) for everyone in the study who consented, for 6 weeks prior to the study and the 3 weeks during the study.

At the end, we used multiple methods for evaluating the results which are all presented below.

\subsection{Participants}
\paragraph{Population} All participants were from one large organization at a global software company. We initially recruited from a list of 10,000 randomly chosen software engineers from around the company. From this group, 337 people completed the survey but only 269 of them agreed to be in the study. From this population we removed people who were not in an allowed country which left us with 228 people in the Intake Population. We used the Intake Population for all analysis of pre-existing beliefs. From this population a total of 106 successfully completed at least 1 diary during the three weeks, completed the final outgoing survey, and were compliant with their treatment, which gives us a Final Population of 106. We used this sub-population for analysing how beliefs had changed and how generative AI was being used. 

The intake population (used for studying pre-existing beliefs) was made up of 188 men, 1 non-bindary / gender diverse person and 4 who prefered not to say; 31 managers and 197 independent contributors; 50 junior engineers, 96 senior engineers and 82 principal or above; 70\% have C as their primary language, 24\% use JavaScript, TypeScript, C\# or another similar language, and 6\% use some other language.

The final population (used for comparison between beginning and end) was made up of 90 men and 16 women; 9 engineering managers and 97 independent contributors; 19 junior engineers; 46 senior engineers; and 41 Principal or above. 53\% of the engineers used C\#; 17\% used C++; 17\% used JavaScript or TypeScript and 13\% reported using some other language including Python, React, Swift, and internal languages.

\subsection{Intake Questions}
All participants were asked a number of incoming questions about their demographics (reported above) and about their current views of GenAI coding tools. They were also asked a number of questions to assess how they feel about themselves and their workplace. The full survey is available online \cite{supplemental-materials}, but here are some examples of what was asked: a Likert scale question on their views of themselves at work, asking for their agreement with statements such as: "I am a highly trained individual", "I possess a unique set of technical skills" and "I cannot be replaced by AI"; a Likert scale question on their views of AI Coding tools, asking their agreement on statements such as "AI coding tools are reliable", "AI coding tools generate correct code" and "I trust AI coding tools; three rating questions to rate their level of energy at work, their average productivity, and how fulfilled they feel about work; an open text question about their opinion of engineers using AI coding tools at work; an open text question on how they think AI coding tools might change work and their place in it.

\subsection{Randomization}

We first divided our candidate pool into those with prior Github Copilot experience and those without prior experience. 
With the group without any prior experience, we randomly sampled them into "treatment" (i.e., uses Github Copilot) and "control" (i.e., continues to not use Github Copilot) groups, so that the effect we observe can be isolated to the first-time introduction of Github Copilot. 
In our sampling, we used block randomization based on three variables: besides gender, we aimed to balance the two groups on their initial perception towards AI coding tools using their agreement to the statements "I like AI coding tools" and "I trust AI coding tools". 
Chi-square test of independence revealed no statistical differences between treatment and control groups. 
We kept the group with prior Github Copilot experience as a third "continuing" group for comparison purposes.
Our initial balancing efforts were somewhat impacted by our exclusion of participants due to study adherence issues. The final breakdown of participants among these three groups can be seen in Table~\ref{table:demographics}.

\begin{table}[h!]
\centering
\small
\caption{Demographics of Randomized Groups. IC refers to independent contributers.}
\begin{tabular}{|l|c|c|c|}
\hline
\rowcolor{graylight}
\textbf{Demographic} & \textbf{Control} & \textbf{Treatment} & \textbf{Continuing} \\
\hline
\rowcolor{grayextralight}
\textbf{Gender} & \textbf{N} & \textbf{N} & \textbf{N} \\
\hline
\hspace{1em} Men & 26 & 25 & 39 \\
\hspace{1em} Women & 3 & 10 & 3\\
\hline
\rowcolor{grayextralight}
\textbf{Level} & \textbf{N} & \textbf{N} & \textbf{N} \\
\hline
\hspace{1em} Level 1 & 1 & 0 & 1 \\
\hspace{1em} Level 2 & 5 & 5 & 7\\
\hspace{1em} Senior & 13 & 19 & 14\\
\hspace{1em} Principal & 10 & 11 & 20\\
\hline
\rowcolor{grayextralight}
\textbf{Role} & \textbf{N} & \textbf{N} & \textbf{N}\\
\hline
\hspace{1em} Managers & 1 & 2 & 6\\
\hspace{1em} IC & 28 & 33 & 36\\
\hline
\end{tabular}
\label{table:demographics}
\end{table}

\subsection{Daily Diaries}
Each day, participants were sent a Teams message to fill out about their experience. All participants were asked a shared set of questions, then there were some unique questions based on group (treatment or control). Each day the diary gathered an average productivity and energy rating (on a scale from 1 for very low to 5 for very high), whether they used GenAI coding tools that day, a description of their day, and how they used GenAI coding tools (if they were in treatment or continuing). If in the control, they were asked how they might have used GenAI coding tools that day if they had access.

\subsection{Closing Survey}
At the conclusion of 3 weeks, all participants were sent a closing survey which repeated all the Likert scale questions from the original survey. It also asked them to self attest if they have been using any generative AI coding tools in the past 3 weeks in order to track compliance.

For those in the treatment group, we asked how satisfied they were with Github Copilot and how they had been using it over the course of the study. We also asked asked how using AI coding tools at work changed the way they worked, and how (if at all) it impacted how they feel about work and their place in it. Lastly we asked the treatment group if they would recommend any changes to AI coding tools. Those in the control group were asked if they would have liked to be using AI coding tools for the last 3 weeks, and why or why not. At the end, all participants were given the opportunity to leave any other thoughts or feedback.

\subsection{Telemetry Collection}
Finally, to be included in the study participants had to give consent for their telemetry to be collected for 6 weeks prior to the study and the three weeks during. This telemetry is already collected by the engineering system used at this company, but participants need to give written consent to allow  researchers access to it. Researchers had access to telemetry such as total code changes, total number of pull requests, time in a development environment and time spent between creation and closing of a pull request.

\section{Results \& Discussion}
\subsection{Incoming values and beliefs about generative AI coding tools}
\subsubsection{Pre-study Github Copilot Usage}
All participants first filled out a survey with their incoming beliefs around generative AI coding tools and Github Copilot as well as about their own skills and place in the workplace. 45\% of participants had heard of Github Copilot at the time, but had not tried it yet. Interestingly, 4\% had not even heard of this tool. We considered 16\% of respondents consistent users, as they either said they have consistently used it for weeks or were a "power user" or early adopter. We asked those who had heard of it but were not using it much to explain why. We found the number one answer to be that they were too busy to explore using a new tool. Since this is a tool to improve developer productivity, the fact that developers felt too busy to use it suggests barriers. Using open coding, we analyzed the open text for why developers had not tried this new tool, and the top reported reasons were:
\begin{itemize}
    \item{Too busy/not enough time (17\% of responses)}
    \item{Don't think it will work well (12\%)}
    \item{Technical blocker (authentication, errors installing, etc) (10\%)}
    \item{Don't feel the need to use it (10\%)}
    \item{Did not know it was available (8\%)}
    \item{Lack of instructions on how to access it (8\%)}
\end{itemize}

Interesting, 6\% of developers reported they were not using it because they did not want to pay for it, even though their company provided it for free. Similarly, a number of people did not realize they had access or were allowed. 
It might be useful for companies to focus on communications to let people know they have access and help them learn how to access and use the tool.
Because coding with generative AI is very different than traditional coding given the non-determinism of these tools, it is important to give developers enough time to learn this new way of working.

\subsubsection{Preexisting Beliefs about Generative AI Tools}
We asked each participant how they felt about generative AI coding tools, such as Github Copilot. Half of the participants had never used these tools, and so their responses were based on what they had \textit{heard} about them rather than on their own experience. Another third had used them a few times. The remaining 16\% were fairly regular users. 

Since some people on the intake survey reported using Github Copilot, we did much of our initial analysis looking at those with experience compared to those without. We defined having Copilot Experience as those who selected "I've used it for a couple weeks", "I use it consistently and have for a few months" and "I'm a power user and\\or was an early adopter" to "How much experience do you have using Github Copilot?". Those who selected "None at all - I haven't heard about it", "I've heard about it but haven't tried it", or "I've installed it and played around a little" are considered not having prior experience. Participants who came in to the study with prior experience were statistically significantly more likely to think AI coding tools are useful in their job (p $<$ 0.05) and like using them (p $<$ 0.05). Specifically, 86\% of those users with prior experience agree that AI coding tools are useful, compared to only 44\% of those without experience. Similarly, 72\% of those who had experience using Github Copilot reported they like using it, compared to only 43\% of those who do not have experience. Considering these tools have been reported as being incredibly helpful (improving productivity by 55\%), it is interesting people are so skeptical before using and find value only once they try it. 

\begin{figure}[h]
    \centering
    \includegraphics[width=0.4\textwidth]{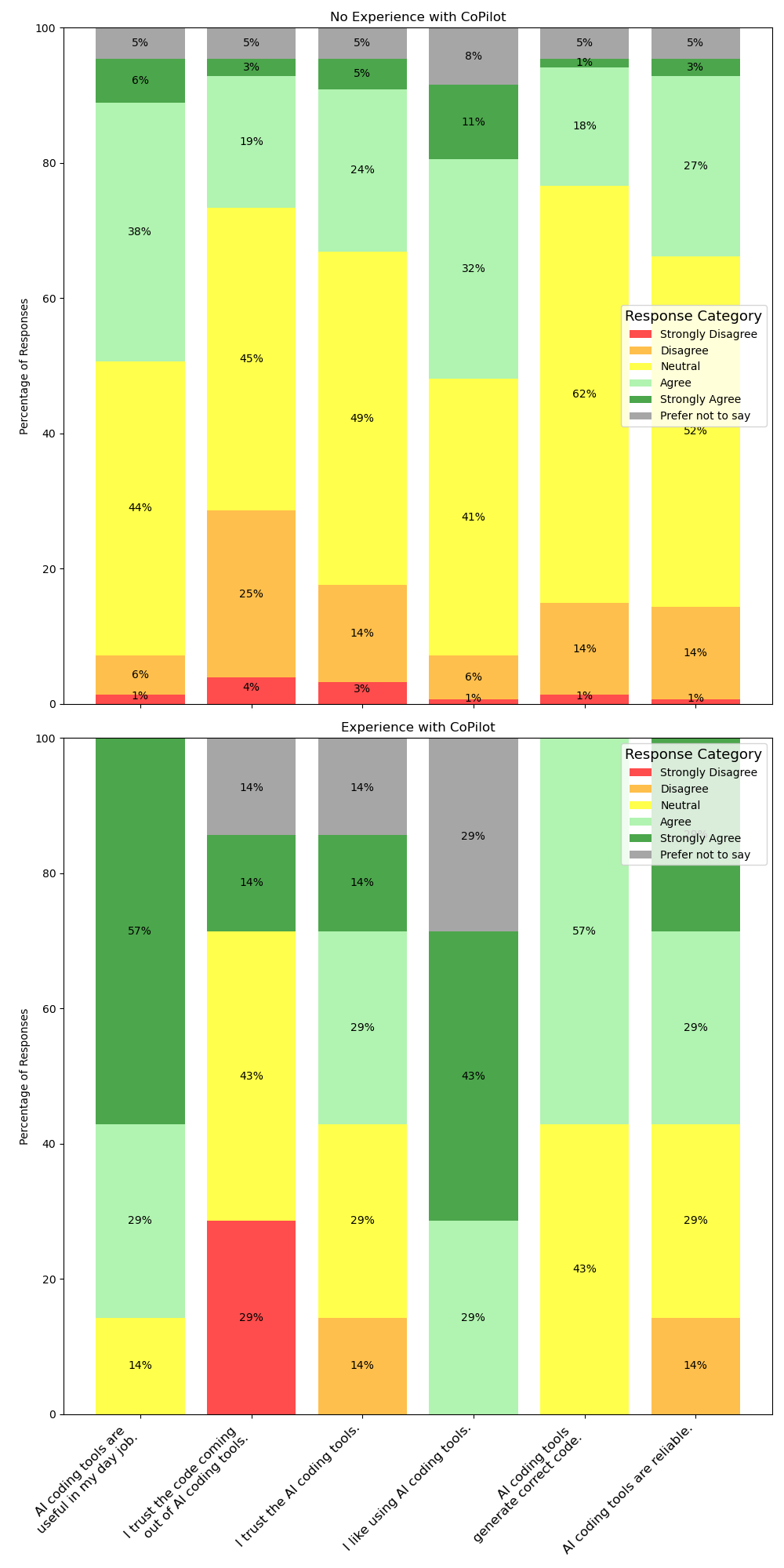}
    \caption{Initial opinions about GenAI coding tools as reported on intake survey. This data is split into those who reported having prior experience and those who did not.}
    \label{fig:incomingByExperience}
\end{figure}

Users with and without prior experience did not differ significantly in their beliefs about whether they trust AI coding tools or the code it outputs, or whether it generates correct code or is reliable. Approximately 20\% of both groups trust the code coming out and think they generate correct code, with the majority of respondents being neutral on both of these questions. Closer to 30\% believe the tools are reliable, but still the majority of both groups are neutral on this as well.  You can see all of both groups responses in Figure \ref{fig:incomingByExperience}.

\label{sec:incomingBeliefs}

\subsubsection{Preexisting beliefs about self and work}
The vast majority of survey respondents felt secure in their ability and the work they do. For example, 88\% agree that they are a highly trained individual, with only 2\% disagreeing. Similarly, 87\% believed they possessed a unique set of skills. Reflecting on the work they do and their work environment, 80\% felt engaged at work, 94\% believed they do work that matters, and 78\% believed they could find a new job fairly easily. Only 65\% agreed that they could not be replaced with AI with the rest being mostly netural, and 10\% believing they could be replaced by AI. At least for now, most people also felt secure in their role, with 78\% of people agreeing. On average, most people felt fairly energized at work with an average score of 4.08 on a scale from 1 and 5. Respondents rated their average productivity slightly lower than their energy at 3.96 on a scale of 1 to 5. They felt similarily fulfilled at work, with an average response of 3.98 on a scale of 1 to 5.

\subsubsection{Beliefs about using AI tools}
The authors asked "What do you think about engineers using AI tools in coding?" to understand how developers perceived others. To analyze the results from this question we used open coding to code the verbatims. The total number of codes reported was greater than the number of people in the study as a person could give a verbatim with multple codes.  Most people (57\%) had positive views, believing that it helped or would help productivity, that it was useful, or were impressed by developers who had already started using these new tools. Some people (20\%) expressed concerns about these tool, or called out that they think they can be useful if a human is in the loop and verifying the output. This is to be expected, as previous studies have found developers have some concerns over the security of these tools \cite{genai_jaworski}. We also know from prior work that GenAI tools do tend to introduce security vulnerabilities (\cite{genai_pearce}). 13\% of engineers were impressed by others using these tools, believing them to be on the forefront of the field. This has been previously seen in the literature where Lamba \textit{et al.} found that early adoptors tended to have a higher social standing \cite{genai_lamba}. A selection of respondents (11\%) believed using these tools was inevitable as they are "the future". Additionally, some (11\%) believed they are useful for certain things, like repetitive work or mundane tasks, but not yet broadly useful when writing complex software.


\begin{tcolorbox}[float=htbp, colframe=blue!50!black, colback=blue!10, title=Research Question Results, fontupper=\small]
\textbf{RQ1: What pre-existing beliefs do developers have about Generative AI coding tools?} \\

\textbf{Summary of Results} 
- Even when encouraged to use these tools in the workplace, 45\% of developers had heard of them but not yet tried them. The most common reasons for this was that they are too busy to try attempting a new tool or that they did not think it would work well.\\
- Developers who have never tried these tools were skeptical. They were signficantly less likely to believe they would be useful or that they would like them, but had similar views of their trustworthiness. \\
- Developers had positive views about others using these tools, often times impressed that they were making use of the latest technology. Some developers (20\%) were concerned about people using these tools at work due to insecurity, tricky bugs and ethical considerations. \\
\end{tcolorbox}

\subsection{Changing beliefs after use}
As stated in the methodology, we split participants into 3 groups based on prior Github Copilot usage - those who had already used Github Copilot and could continue to use it (Continuing); those who had not being using it and were asked to continue not using it (Control); and those who had not been using it but were asked to use it for the next 3 weeks (Treatment). After three weeks of using (or not) GenAI, all participants repeated the questions on beliefs around themselves, their work and generative AI tools to look for any changes. We then analyzed the before and after beliefs for all participants.

We looked at the average response level during the intake survey and final survey for every Likert question, and compared them using a paired t-test. For the statement "I like using AI coding tools," the average rating rose from 2.72 before to 3.61 after, with $p < 0.0001$. In addition, there was statistically significant increase in response to the statement "AI coding tools are useful" with a rise from 2.93 to 3.51 and $p = 0.001$. Lastly, there was an increase for the treatment group in response to "AI coding tools generate correct code", from 2.71 to 3.0, but this was only borderline statistically significant ($p=0.051$). Interestingly we saw a statistically significant rise in both groups (treatment and control) for "I possess a unique set of technicals skills". The average treatment group response rose from 4.22 to 4.46 ($p=0.05$) and the control group rose from 4.23 to 4.46 ($p=0.02$). It is unclear why both groups increased in their belief that they have a unqiue set of skills. One hypothesis is that this study was conducted at a time where new AI announcments were being made regularly, and they dominated the news cycle. There was much talk of whether these tools could replace developers. We postulate that developers were growing more sure of themselves as they saw more of what AI could, and more importantly \textit{could not} do. \\

There were no statistically significant changes in other variables, including how developers felt at work or in their level of trust for GenAI coding tools. The increase in liking after using reflects what has previously been seen in the literature where use of GenAI tools drives adoption or people enjoyed using the tools after being given access \cite{genai_chatterjee}, \cite{genai_sergeyuk}, \cite{genai_pereira}. 

Correlation analysis found that being satisfied with AI was slightly more correlated with a sense of liking AI and finding it useful than the actual trustworthiness of AI and the code it produces (r = 0.691 vs r = 0.606). Here, "r" refers to the correlation estimate, which ranges from 0.0 to 1.00, with higher numbers indicating stronger correlation between two variables. A regression analysis with AI satisfaction as the dependent variable confirmed that AI utility is a stronger predictor of satisfaction than AI trustworthiness (although both sentiments are highly predictive). This is consistent with the increase of liking GenAI after using it, but not increasing in trust of GenAI. It seems developers can enjoy using the tool, but also understand the need to validate what comes out of it and be critical of output. This helps them be more prodcutive overall, while reminding them that they have unique value as a human contributor at the company. Indeed, in the diary verbatims, some people mentioned that when GenAI tools gave them incorrect answers, they still enjoyed that it helped them think about the problem in a new way: "AI tools were certainly giving lots of suggestions, some led me down the wrong path, but regardless, was good at spinning up a different way of thinking" (P65). Since concerns in the intake suvey included fear of being replaced by AI, it is very positive to see that actually using these tools helps developers feel more comfortable with them and confident in themselves.   

We also conducted a comparison of participants' pre- and post-study beliefs regarding the use of GenAI by others. Overall, beliefs about others using the tool remained relatively stable after the study. However, a notable increase was observed in the number of developers who perceive GenAI tools as inevitable and representative of "the future." For instance, as P66 stated in the exit survey, "I think it is the future of coding. Engineers not using AI for coding are missing out." Additionally, there was a slight increase in the number of developers emphasizing the need for verification of code produced by GenAI tools. Given the importance of striking an appropriate balance in the reliance on GenAI--neither over-relying nor under-relying--this shift in perception is a positive development that may contribute to more effective use of these tools in practice.

\subsubsection{Telemetry Results}
Six weeks of telemetry from before the study were compared with the three weeks of telemetry from during the study. We chose to use a difference-in-difference (DiD) approach because we have developers' telemetry collected at multiple time points (before and after the start of the study), and we want to account for any underlying trends related to the time point that might affect the outcome.

For each metric, we computed the average values across all three assignment groups (Treatment, Control, and Continuing) over the entire period of data collection. Notably, differences between the Treatment and Control groups were already apparent across several metrics prior to the intervention. For instance, interaction time consistently exhibited higher values for participants in the Treatment group compared to those in the Control group. These observed differences supported the validation of a key assumption underlying the difference-in-difference (DiD) approach: the parallel trends assumption. This assumption is critical, as it posits that, in the absence of treatment, the differences between the treatment and control groups would have remained constant over time.

We did observe medium-to-large differences for most of the metrics, but we did not have any statistically significant difference-in-difference results. This means that having access to the Gen AI tool made no statistically significant difference to developer telemetry metrics. Table \ref{table:did_results} shows the DiD coefficients, p-values, and power for each of the metrics we examined.

\begin{table}[h!]
\centering
\caption{DiD Coefficients, P-Values, \\
and Power for Metrics}
\begin{tabular}{|l|c|c|c|}
\hline
\textbf{Metric} & \textbf{Diff-in-Diff} & \textbf{P-Value} & \textbf{Power} \\
\hline
CodeChanges & -89.9 & 0.5 & 0.06 \\ \hline
PullRequests & 0.1 & 0.9 & 1.0 \\ \hline
PullRequestHours & 2.6 & 0.8 & 0.2 \\ \hline
DevelopmentMins & -21.1 & 0.6 & 1.0 \\ \hline
EmailMins & -7.5 & 0.6 & 1.0 \\ \hline
BuildMins & -52.0 & 0.8 & 1.0 \\ 
\hline
\end{tabular}
\label{table:did_results}
\end{table}

There could be multiple reseasons for the lack of statistically significant results. For one, perhaps the study was too short for learning and getting proficient with a new tool. In fact, at the time we were unsure how long it would take for new users to see an impact from using GenAI tools. Now research suggests that it might take 11 weeks of using GenAI tools (in this particular study, M365 Copilot, a similar productivity Copilot from Microsoft) for 11 minutes a day to hit the "tipping point" where users report improvements in productivity \cite{tippingpoint}, \cite{prodReport}. This may have resulted in an overall low effect size -- the actual effect of the treatment might be too small to detect with this amount of data. Future researchers should consider looking at longer duration, perhaps 6 months to a year, to see if there are significant changes. In addition, the diaries showed that many engineers in the treatment group did not get to code as much as they would like over the three weeks (due to things like sprint planning, being oncall, vacation, etc). These confounding factors could not be accounted for in this difference-in-difference model. Also, only 75\% of both groups (treatment and control) were fully compliant - with a quarter of the control group reporting they had tried generative AI tools during the course of the study, and a quarter of the treatment group saying they never used generative AI tools. This reduced our population for study, potentially making our group too small to see changes in such specific metrics.


\begin{tcolorbox}[float=htbp, colframe=blue!50!black, colback=blue!10, title=Research Question Results, fontupper=\small]
\textbf{RQ2: How do pre-existing values and beliefs change after regular use of GenerativeAI code tools in daily work?} \\

\textbf{Summary of Results} \\
- The use of Github Copilot (and other GenAI tools) in day to day coding work statisticaly significantly increased developer's beliefs that these tools are useful and enjoyable.  \\
- No significantly significant changes were seen in telemetery measurements of the developers, including lines of code written, PRs completed, or time spent coding. \\
- Changes in beliefs around utility and enjoyment were more important than trust when it comes to being satisfied with AI coding tools. Even though participants sometimes found them incorrect, some still enjoyed using them and even used the tool's incorrect output to help them think differently about coding.
\end{tcolorbox}

\subsection{Use Cases}
\label{sec:UseCases}

\subsubsection{Diary Results}
Participants were contacted daily at the end of the work day to fill out a short diary about what work they did that day, how they used AI (or how they would), and if they used AI, how helpful it was. We received 1,235 dairies in total, with participants submitting anywhere from 1 diary over the 3 weeks to 18 diaries. We received a mean of 8.37 diaries per person, with a standard deviation of 4.819. A Kruskal-Willas test found that the distribution of diary response numbers was similar between the groups ($p = 0.412$). These diaries included a wealth of information on how developers were using GenAI in their real work day, and we include many verbatims throughout this paper in order to share the rich qualitative data gathered in the diary. 

GenAI use was reported in 56\% of diaries in the Treatment and Continuing group (this question was not asked in the control, since they were not allowed to use GenerativeAI). GenAI coding tools weren't used 100\% of the time because developers do a lot more work than just coding. In fact, previous studies show a good day is one in which developers are able to code for 18\% of their time (vs 11\% on a bad day) \cite{goodday} while another found developers spend 21\% of their time coding, with time also allocated to emails (14.5\%), collaborating (24.4\%), and looking at work related websites (11.4\%) \cite{meyer}. In the diary, people in the treatment group did often report they were not able to use GenAI on a given day because they were in meetings, working in a language or code base that didn't support it or in a design phase instead of coding phase. This would mean they had less time to try Github Copilot, and less time to get proficient at it. Some examples of this from the daily diaries:

\interviewquote{253}{ Did not get a chance to do any coding today (which isn't necessarily unusual, but is frustrating for me none-the-less).}

\interviewquote{65}{Catching up on emails, no coding}

\interviewquote{289}{was mostly editing yaml and json config files for a new build type so AI is not helpful for that (gets more wrong than right)}

Interestingly, several participants reported finding generative AI tools useful even on non-coding days, utilizing them for tasks such as debugging or as a tool for brainstorming design ideas. For specific examples, refer to Section \ref{sec:UseCases}. This observation aligns with a broader pattern in our findings, where some developers expressed frustration at being unable to use GenAI for certain tasks, such as learning a new codebase or setting breakpoints, while others found innovative ways to leverage GenAI for precisely those use cases. In some cases, the same user expressed frustration, then discovered how to use GenAI for that use case later on. This recurring theme underscores the importance of understanding the full range of applications for generative AI tools, as this knowledge is crucial for realizing their potential value. You can see this pattern in the verbatims below:

\interviewquote{259}{I am trying to learn a new codebase. I wish AI could help me identify the best examples in the codebase to learn from.}
\interviewquote{259}{I tried learning a new language (Graphviz) today in VSCode and Copilot made several helpful suggestions as I was getting the hang of the syntax.}

Here are sample verbatims showing the same pattern, this time with some developers using GenAI to help them debug code, and others lamenting that GenAI tools can't help with debugging:\\
\interviewquote{262}{Given what I'm working on, AI tools weren't super useful. I'm doing a lot of debugging vs writing new code.}
\interviewquote{197}{Debugging an issue with windbg... I used copilot to answer several questions about commands, formatting.}

Some of what was reported in the diary validated the incoming beliefs developers reported. For example, the most common cateogry of answers when asked what they think of engineers using GenAI coding tools was that it would improve their productivity. People \textit{did} report an improvement in productivity from using these tools, such as \interviewquote{238}{Multiple code implementation suggestions by copilot chat in VisualStudio made productivity 50\% more}. The second most common incoming thought when reflecting on engineers using these tools were concerns around the use, that they might introduce bugs or degrade the code. This is reported on more below, in Challenges. As we've reported earlier, using the tools also helped developers want to continue using them, as reflected by this daily diary statement: \interviewquote{197}{Really happy about the AI tools today, I'm more willing to try to use the tools to ask questions that I wouldn't have even attempted a few days ago.} While some respondents felt that the time spent to verify an answer meant they weren't seeing any productivity gains, others found that while AI wasn't always correct, it helped them think in differnet ways or learn something new. 

A common use case observed in the daily diaries was that of helping write boilerplate, repetitive code, docs or comments. The notion that AI could assist with repetitive tasks and boilerplate code was one of the incoming beliefs, and this was evidenced in diaries. Many reflected on how it saved them time by doing repetitive work for this, for example:

\interviewquote{155}{I was SHOCKED at how well it wrote docs, and even copied the formatting exactly. I'll happily write docs like this forever.}
\interviewquote{129}{I liked having help with adding comments. I was impressed that I only had to update a few things in the comments it generated, and I didn't have to spend time thinking about the best way to phrase an explanation of what the code did (I sometimes struggle with translating code to human language, so it was nice to have a jumpstart)}

A very common use case reported in the diaries was one that had not been mentioned much in the intake - that of using it as a replacement for websearch. Many people reported using generative AI to learn new things about an application, search for information, and find answers to niche questions. For example:
\interviewquote{197}{pretty neat!  I was able to get answers about things that aren't very easy to find in online searches.}
\interviewquote{119}{New to writing JEST tests. Instead of having to dig through documentation and stackoverflow, I had a conversation with AI to build my first tests. It was much quicker and continuous as a conversation rather than a splintered web search}

Lastly, developers found that these tools can be very useful but need to be used in the right way. Some users found patterns that worked to maximize the utility of Copilot, while others found it seemed to improve on its own overtime. 
\interviewquote{27}{Today showed that AI tools can be useful, but it really needs to be done in an iterative way: Solve a specific small problem or take an initial solution and improve on it. It also often needs to be forcibly told what to do.}
\interviewquote{259}{Today the AI gave me some amazingly accurate suggestions. Maybe it's just because I've been working on this code for several days now, but there were some multi-line completions that were exactly what I needed.}

\subsubsection{Challenges}
While many people reported Copilot improving productivity and being enjoyable to work with, people still had challenges and concerns using it. The top issues reported were: code that appears correct but actually is not (an incoming concern that was validated in the diaries); the need to validate generated code negating the produductivity gains; and Github Copilot not working in a variety of contexts, codebases and languages.

Code that \textit{appears} correct but is not is a serious concern, as many engineers might not notice a subtle bug. Some respondents initially had a fear that junior engineers should not use tools like this as they might not have the experience to notice syntatically correct but technically wrong code. In the diaries, we saw examples of these kinds of tricky mistakes. For example:
\interviewquote{154}{Noticing the suggestions aren't as smart as they look. They are deceivingly smart since they'll get very close but then not account for some basic thing, which is often worse than appearing bad}
\interviewquote{129}{CoPilot is really neat when it works (and I did get super excited when it popped up a whole function body that looked mostly okay on first glance, since it mostly just gives me line-by-line autocompletion), but it's kind of frustrating when it doesn't.  Many of the property names it used were similar to the actual names, but not quite right (or worse, valid names that weren't the property I was looking for), so I didn't notice all of the issues immediately, and I find it harder to quickly find mis-matches like that when I'm looking at already written code vs. when I'm writing it.  My excitement at having a big chunk of code written for me was replaced with a sigh.}

It is known that generative AI models can hallucinate, producing incorrect output. Since all generative AI content should be validated, some engineers found themselves wondering if it really saved them anytime at all. For instance, one respondent shared:
\interviewquote{129}{I was frustrated that when adding to the config file it did not give me a new guid for my new entries (it pulled one from the next entry in the file) unlike the last time I used it to do so. Now I don't know if I can trust it, which means it won't really save me time.}

Lastly, Copilot does not work on all languages equally. It works especially well for Python, JavaScript, TypeScript, Ruby, Go, C\# and C++ \cite{githubWebsite}, but people complained it did not work well with less common languages and file types, making it less useful overall.

\interviewquote{289}{I was working with Android bp/mk files today.  It seems copilot doesn't really know those, which makes sense (they are quite special). [...]  Copilot might have tried to "help" me, but it was just  guessing and it really isn't good at that with these types of files.}
\interviewquote{31}{Another big failure is .json files. While C\# files suggestion makes sense, suggestion attempted on config files are never relevant.}

While the lack of support for certain programming languages is expected to improve over time, as generative AI systems become more adept at handling niche languages, other challenges, such as hallucinations and the need for thorough validation, are inherent to the nature of generative AI. To fully realize the productivity benefits of these tools, it is essential to understand the contexts in which they are most effective and to recognize their limitations. We have entered a new paradigm of software development, where code can be generated more quickly than ever before, but requires a greater degree of critical analysis. This "faster but with more mistakes" approach diverges from traditional practices, necessitating rapid evaluation and constant trade-offs to determine whether the time spent validating AI-generated code outweighs the efficiency gains. We observed that developers experienced significant productivity benefits when using generative AI for low-risk tasks, such as generating documentation, comments, and repetitive code, as well as for creative processes like idea generation. However, in cases where errors could be more difficult to detect, such as the auto-generation of GUIDs, developers did not report similar productivity gains due to the cognitive effort required to validate the code. Prior research in the field of automation highlights that humans are generally poor at performing monitoring tasks \cite{mackworth}, suggesting that reviewing AI-generated code will likely remain a persistent challenge.


\begin{tcolorbox}[float=htbp, colframe=blue!50!black, colback=blue!10, title=Research Question Results, fontupper=\small]
\textbf{RQ3: What are the actual use cases and challenges of using GenerativeAI in real work?} \\

\textbf{Summary of Results} \\
- Developers used GenAI in a multitude of ways, some they predicted (like for boilerplate and repetitive code) and some they did not (as a replacement for websearch).
- Developers exhibited variability in their discovery of use cases for generative AI tools. A particular use case that one developer finds ineffective or problematic may be the same use case in which another developer reports significant productivity gains.
- GenAI challenges are largely to be anticipated, but signal a requirement for a new way of working, with more emphasis placed on code validation. 
- Challenges associated with generative AI can be mitigated by selectively applying it to low-risk tasks, such as generating comments, boilerplate code, and documentation, where the impact of errors is more likely to be minimal.
\end{tcolorbox}

\subsection{Impact of using Generative AI Coding Tools}
After the three weeks of study, participants submitted a final survey where they were asked the same questions as incoming (described above) but also asked "How has using AI coding tools at work changed the way you worked?". Using open coding, we coded the 94 open text responses we received from this question. The most common response was that using AI coding tools saved time and/or improved productivity. The second most common response was that it reduced the time spent (or in some cases removed the time spent) writing boilerplate code and repetitive code. Multiple users also mentioned that it kept them in the IDE and largely replaced web search; that it helped them learn a new language; and that it made them more likely to write unit tests and documentation. Interestingly, you'll see below that even when the code coming from AI tools isn't correct, for some it sparks better ideas and still helps. Some verabtims for these categories:
\interviewquote{106}{It improved the the speed of coding and made it seamless. Entry barrier for a new language is very small now.}
\interviewquote{212}{Helped with new languages. Faster development. Did not have to reach out to search platforms or websites often for help.}
\interviewquote{77}{It has given me the possibility to efficiently do the laborous work with ease. Generate basic boiler plate etc. which helps in faster development and testing. I am also leveraging it to learn new languages like Python.}
\interviewquote{65}{Saves me a lot of time. I often use suggestions, although not always correct, they add a spark that sometimes leads to better ideas.}

There were 99 verbatims about how using AI changed their work. Some verbatims included multiple codes, making a total of 129 counted codes. In total, 84\% of the codes were positive, describing changes like the above. However, 16\% were negative and included responses such as "no change", it slowing them down, or not being useful for their current role. Some examples of these are:
\interviewquote{157}{Unfortunately, it hasn't [changed]. I was really excited to try it out but throughout the course of the study I just didn't really see it having any impact. I didn't really go out of my normal patterns, but there also weren't really any usage instructions or examples on the web of how to best use them.}
\interviewquote{259}{There were a few cases (less than 5 during the 3 weeks) that it prompted me with the answer and I didn't need to look up the syntax or keyword. Otherwise, the only way it changed how I worked was it saved me some typing (which I still had to proofread)}
\interviewquote{16}{It slows me down and gets in the way. It is constantly interrupting with irrelevant or wrong autosuggestions.}

It was very interesting to see some people changed how they worked to make code better for AI. For example, one user said \interviewquote{131}{now I try to make my code more explicit to help the AI give me suggestions}.

In addition to asking how these tools changed how they work, we also asked "How has using AI coding tools impacted how you feel about your work and your work place?" and again used open coding to code the 88 verbatim responses. For this question, 38\% said there was \textbf{no change}, or \textbf{no change \textit{yet}}. For those that noticed a change, the most reported change was that it \textbf{increased work satisfaction or just felt good}. After this, the most common change was that work was \textbf{less boring} since boilerplate and repetitive code was taken off their plates. People also reflected on the place these tools would have in the workplace. While some people said it made them feel less threatened and glad that humans were still writing code, others expressed a fear of being replaced soon or needing fewer engineers. Others recognized that work is changing and that they will need to adapt. Some verbatims that express these sentiments can be found below.
\interviewquote{129}{I'm definitely sure I can't be replaced by an AI coding tool (at least not in their current state!), but I do think it will be interesting to see how they grow.  I am more aware of how many poorly named variables \& property names we have in our codebase now, but I'm also more incentivized to ensure I don't add to the problem.}
\interviewquote{284}{More excited to code than ever before.}
\interviewquote{125}{I feel less stressed now. Its like I have a 24x7 companion who will help me solve my problems.}
\interviewquote{164}{Great to have access to this tooling, so it has certainly greatly improved how I feel about my place of work.}
\interviewquote{11}{Github Copilot makes some of the work easier. It alsos hows fewer devs will be needed in the near future.}


\begin{tcolorbox}[float=htbp, colframe=blue!50!black, colback=blue!10, title=Research Question Results, fontupper=\small]
\textbf{RQ4: What is the impact of using GenerativeAI tools in developer's daily work, and how does it change their and their place in it?} \\

\textbf{Summary of Results} \\
- Most developers found Github Copilot made them more productive with 84\% reporting some kind of positive change in how they work. Some reported it largely replacing web search engines and many reported they now write less boilerplate or repetitive code and can focus on more fun/challenging work.\\
- 66\% of developers reported a change in how they feel about work after using these tools. Some changes included being more excited and optimistic about their work while others reported feeling a need to upskill and not be left behind.
\end{tcolorbox}

\section{Limitations}

\added{In this section, we discuss limitations of this work.

This research relied on self-reported data from surveys and diaries, which introduces potential biases such as social desirability and recall bias \cite{podsakoff2003common}. However, self-reported data is valuable in organizational research, especially when objective measures are impractical \cite{conway2010reviewers}. While we collected quantitative telemetry to observe how GenAI influenced work, our primary focus was on how GenAI affected personal views about work and one's role. Given this interest, self-reported data was the most viable approach, despite its acknowledged limitations.

The questions in the survey we deployed were original and not previously validated since, at the time of the initial study (summer 2023), we did not believe there were any randomized controlled trials of Github Copilot in the workplace (as opposed to in a lab). 
While this meant we could tailor our questions to our specific research, it may impact the generalizability of our findings. 
We share these questions in supplemental materials~\cite{supplemental-materials} and invite subsequent studies to validate and refine them.
 }

Finally, this work was conducted at a single company, and some believe single company results will not be generalizable. However, Flyvbjerg shows individual case studies that led to discoveries in physics, economics, and social science~\cite{flyvbjerg2006five}.
This is not a criticism of research done on large and different groups in the population. Both types of research are necessary and essential to build up our body of knowledge \cite{basili1999building}. In addition, while this work was done at one company, this company has teams all over the world, working in different languages and cultures, which makes it less homogeneous than the average single company case study.


\section{Conclusion}
In the final survey, one person expressed a concern for us moving in this direction, and gave a challenge to seriously consider what we are doing with these new tools. Participant 245 said "Any time you are using code you do not understand, you are potentially introducing vulnerabilities. We are slowly chipping away at our understanding of each the parts, and with that comes a slow degradation of our understanding of the whole. These are dangerous waters we are sailing into unprompted, and we should strongly consider what lies at the end of this path before fully setting our sails." This study was an attempt to do just that - to look at the impact these tools might be having on our developers, and not just on their productivity, but their beliefs about themselves and their role in the workplace. To our knowledge, this study was one of the earliest randomized controlled trials of generative AI tools in the real workplace, over a period of time via diary study. We also believe this study to be one of the first looking at how generative AI work tools can impact beliefs and values at work. 

We found that the impact of generative AI tools at work is nuanced, with positive and negative effects, although overall more positive effects were reported. The use of these tools significantly increased developers beliefs that the tools were useful and enjoyable, as seen in Likert question changes and diary study reflections. Most engineers found them to be helpful with repetitive and boilerplate code, and others used it for even more complex tasks, like learning new languages, expanding their thinking, debugging and improving their code bases. These tools positively changed how 84\% of engineers worked, and changed (whether positively or negatively) how 66\% of developers feel about their work and their place at work. For some, they are more excited than ever for the challenge of their role and new ways of working; for others, they see this as a time of rapid change and feel a need to adapt or be left behind. 

While the impacts of generative AI vary greatly based on coding language, task type and how people approach them, one thing seems clear: generative AI tools are changing work, mostly for the better, but the full impact is yet to be seen and needs to continue to be explored. As Participant 72 put it on the final survey: "I think that as with any tools, they have a place. I think that the tools effectiveness is directly related to how educated the developers are in using them. As such, I believe a significant investment should be made in engaging case studies on how to practically leverage these AI tools in day-to-day scenarios". We recommend that organizations seeking to integrate generative AI coding tools into their workflows prioritize educating developers on best practices for their effective use, including the critical task of validating AI-generated code. Additionally, it is essential to update testing infrastructures to accommodate the unique characteristics of AI-assisted development. When applied appropriately, generative AI tools have the potential to enhance developers' engagement, foster greater enjoyment in their work, and instill a sense of being at the forefront of technological innovation.

%
\begin{acks}
We thank all participants in survey and diary study for their insightful comments and committment to this longer term study.
We would also like to thank \added{Josh Pollock, David Speirs, Barrett Amos and Cory Hilke} for their support and feedback.
The ethics for this study were reviewed and approved by the Microsoft Research Institutional Review Board (MSRIRB), which is an IRB federally registered with the United States Department of Health \& Human Services. 
\end{acks}

%
\bibliographystyle{ACM-Reference-Format}
\bibliography{ref}

%

\end{document}